\documentclass[aps,prc,twocolumn,groupedaddress,a4paper]{revtex4-1}
\usepackage{amsmath}
\usepackage{amssymb}
\usepackage{graphicx}
\usepackage{textcomp}
\usepackage{braket}
\usepackage{tabularx}
\usepackage{booktabs, hhline}
\usepackage[dvipsnames]{xcolor}

\newcommand {\la} {\langle}\newcommand {\ra} {\rangle}
\newcommand {\beq} {\begin{eqnarray}}
\newcommand {\eeqn} [1] {\label{#1} \end{eqnarray}}
\newcommand {\eol} {\nonumber \\}
\newcommand {\ve} [1] {\mbox{\boldmath $#1$}}

\begin{document}

\title{Effects of an induced three-body force in the incident channel of $(d,p)$ reactions}

\author{M. J. Dinmore}
\affiliation{Department of Physics, Faculty of Engineering and Physical
Sciences, University of Surrey, Guildford, Surrey, GU2 7XH, United Kingdom}
\author{N. K. Timofeyuk}
\affiliation{Department of Physics, Faculty of Engineering and Physical
Sciences, University of Surrey, Guildford, Surrey, GU2 7XH, United Kingdom}
\author{J. S. Al-Khalili}
\affiliation{Department of Physics, Faculty of Engineering and Physical
Sciences, University of Surrey, Guildford, Surrey, GU2 7XH, United Kingdom}
\author{R. C. Johnson}
\affiliation{Department of Physics, Faculty of Engineering and Physical
Sciences, University of Surrey, Guildford, Surrey, GU2 7XH, United Kingdom}

\date{\today}

\begin{abstract}
A widely accepted practice for treating deuteron breakup in $A(d,p)B$ reactions relies on solving a three-body $A+n+p$ Schr\"odinger equation with pairwise $A$-$n$, $A$-$p$ and $n$-$p$ interactions. However, it was shown in [Phys. Rev. C \textbf{89}, 024605 (2014)] that projection of the many-body $A+2$ wave function into the three-body $A+n+p$ channel results in a complicated three-body operator that cannot be reduced to a sum of pairwise potentials. It contains explicit contributions from terms that include interactions between the neutron and proton via excitation of the target $A$. Such terms are normally neglected. We estimate the first order contribution of these induced three-body terms and show that applying the adiabatic approximation to solving the $A+n+p$ model results in a simple modification of the two-body nucleon optical potentials. We illustrate the role of these terms for the case of $^{40}$Ca($d,p$)$^{41}$Ca transfer reactions at incident deuteron energies of 11.8, 20 and 56 MeV, using several parameterisations of nonlocal optical potentials.
\end{abstract}
\maketitle

\section{Introduction.}
Transfer reactions provide a useful way of probing the structure of nuclei because differential cross sections show features that are sensitive to the shell structure of the target and residual nuclei. This is attractive for experimentalists looking to probe the structure of a nucleus, and $(d,p)$ reactions are a popular choice at radioactive beam facilities for determining the spectroscopic strength of single particle states for nuclei beyond stability. One theory available for analysis of $(d,p)$ reactions is the Adiabatic Distorted-Waves Approximation (ADWA) \cite{JT}. The ADWA accounts for deuteron break up effects through a three-body, $n+p+A$, description of the deuteron channel,  represented by the wave function $\Psi^{(+)}_d$ in the $(d,p)$ transition amplitude  
\beq
T_{(d,p)} = \sqrt{S} \braket{\chi^{(-)}_p \phi_{n} |V_{np}|\Psi^{(+)}_{d}},
\eeqn{tdp}
where $\chi^{(-)}_{p}$ is the proton channel wave function, $\phi_{n}$ is the normalised bound-state wave function of the transferred neutron in the final state (more generally, the normalized neutron overlap function) and $S$ its spectroscopic factor.

It is standard to calculate $\Psi_{d}^{(+)}$ using a three-body model consisting of the three $V_{np}$, $V_{pA}$ and $V_{nA}$ pairwise interaction potentials between the $n$-$p$, $p$-$A$ and $n$-$A$ pairs, respectively. A two-body $p$-$A$ or $n$-$A$ scattering model based on the Feshbach projection operator technique \cite{Feshbach} involves an energy-dependent nonlocal optical potential, implicitly accounting for coupling to excited target states. Applying the Feshbach approach to the three-body $n+p+A$ channel has two implications \cite{Joh14}:
\begin{enumerate}
    \item The $n$-$A$ ($p$-$A$) optical potential in the $n+p+A$ system depends on the proton (neutron) dynamical variables and on the $n$-$p$ interaction.
    \item In addition to the pairwise interactions, new terms arise which  correspond to an effective  interaction between the neutron and proton in the deuteron, via excitation of the target $A$, to all orders. These create a three-body interaction that has both diffractive and absorptive parts.
\end{enumerate}
The dependence of the  optical potential of each $N$-$A$ subsystem within the $n+p+A$ three-body system on the position of the other nucleon can be averaged out if the wave function is expanded over some appropriate $n$-$p$ basis states. It was shown in \cite{Joh14} that choosing the Weinberg expansion \cite{JT} and retaining leading order terms only (which corresponds to the ADWA) results in an effective energy at which the $N$-$A$ potential should be used in $(d,p)$ calculations. However, applications of this idea in \cite{Joh14,Wal16} treated only the pairwise interactions in the $n+p+A$ system, ignoring the additional three-body terms.


One way to deal with these induced three-body (I3B) terms is to explicitly include excited target states in the reaction model. This has, for example, been done within the CDCC \cite{Gom15,Gom17} and Faddeev \cite{Del15,Del16,Del17} approaches. However, these calculations include explicitly only a fraction of the model space needed to fully account for all the absorption known to be needed in the nucleon optical potentials. In this paper, we point out that the ADWA allows us to estimate the contribution of the effective I3B interaction, to first-order, from all target excited states. We explain this procedure in Section II showing that it results in a simple modification of the distorting potential in the deuteron channel and in Section III we describe their connection to the dynamical part of the nonlocal dispersive optical model (NLDOM). In Section VI we discuss nonlocal scattering inputs for the ADWA with NLDOM. In Section V we use the NLDOM potential with the required alterations and we present our numerical calculations using the $^{40}$Ca$(d,p)^{41}$Ca reaction as an example. For comparison, we also present calculations with two other nonlocal optical potentials. Discussion and conclusions are presented in Section VI.

\section{Effective interactions in the ADWA deuteron channel.}
For the case of two-body scattering of a nucleon $N$ on a complex target $A$, Feshbach shows, using the projection operators $P_A$ and $Q_A$ which project onto the ground and excited states of nucleus $A$ respectively, the total many-body wave scattering function is split into two parts, $\Psi_P=P_A\ket{\Psi_N^{(+)}}$ and $\Psi_Q=Q_A\ket{\Psi_N^{(+)}}$. Feshbach then shows that $\Psi_P$ appears in a modified Schr\"odinger equation \cite{Feshbach2}
\beq
(E-H_{PP})\Psi_P=0,
\eeqn{FeshSchr}
and this is governed by an effective Hamiltonian
\beq
H_{PP}=T_{NA}+V^{opt}_{NA}(E_N).
\eeqn{FeshHam}

In the case of two-body scattering of a nucleon $N$ Feshbach \cite{Feshbach} demonstrates that for a nucleon with kinetic energy $E_N$ the optical potential $V^{opt}_{NA}$ takes the form
\begin{equation}
V^{opt}_{NA}(E_N)=\braket{\phi_A|U^{opt}_{NA}|\phi_A},
\label{feshVopt}
\end{equation}
where $U^{opt}_{NA}$ is the optical model operator
\begin{equation}
U^{opt}_{NA}=v_{NA}+v_{NA}Q_A\dfrac{1}{e_N-Q_Av_{NA}Q_A}Q_Av_{NA}.
\label{feshUNA}
\end{equation}
Here $v_{NA}$ is the sum of interactions of the nucleon $N$ with all nucleons in the target $A$ 
and the energy denominator $e_N$ is given by $e_N=E_N+i0-T_{NA}-(H_A-E_A)$, where $T_{NA}$ is the $N$-$A$ relative kinetic energy, $H_{A}$ and $E_{A}$ are the internal Hamiltonian and the ground state energy of the target $A$, respectively.

In previous work \cite{Joh14} this idea of the optical potential was extended for a three-body case. The three-body wave function in this case can be considered as the projection $\Psi_P$ of the full many-body wave function, $\Psi_d^{(+)}$, of the $n+p+A$ system onto the ground state of the target $A$. This projection is governed by the effective Hamiltonian
\begin{equation}
H_{\text{eff}}=T_{3}+V_{np}+\braket{\phi_{A}|U|\phi_{A}},
\end{equation}
where $T_{3}$ is the three-body kinetic energy operator and $V_{np}$ is a short range $n$-$p$ interaction. The operator $U$ is an operator in all $A+2$ coordinates of $n$, $p$ and $A$. The final term implies integration over the target nucleus coordinates to leave an operator in three-body coordinates only. 
We can express $U$, which accounts for the excitation of target nucleus degrees of freedom on the target ground state projection of the scattering wave function, in terms of operators $U_{pA}$ and $U_{nA}$ that define excitations of $A$ by $n$ and $p$ separately \cite{Joh14}. Up to second order in $U_{NA}$ these terms are
\begin{equation}
\begin{aligned}
U^{(0)}&=U_{pA}+U_{nA},\\
U^{(1)}&=U_{nA}\dfrac{Q_{A}}{e}U_{pA}+U_{pA}\dfrac{Q_{A}}{e}U_{nA}, \label{u1}\\
\end{aligned}
\end{equation}
where
\beq
U_{NA}&=&v_{NA}+v_{NA}\dfrac{Q_{A}}{e-Q_{A}v_{NA}Q_{A}}v_{NA}.
\eeqn{UNNA}
However, because of the definition of $e$, given by
\begin{equation}
e=E_{3}+i0-T_{3}-V_{np}-(H_{A}-E_{A}),\label{e}
\end{equation}
the $U_{nA}$ is an operator in both $n$ and $p$ coordinates despite including the $n$-$A$ interaction only. The same is true of $U_{pA}$. In Eq.(\ref{e}) $E_{3}$ is the three-body energy, related to the incident centre-of-mass kinetic energy, $E_{d}$, and deuteron binding energy, $\epsilon_{d}$, by $E_{3}=E_{d}-\epsilon_{d}$. The $U_{NA}$ is not equal to the Feshbach optical operator that describes the two-body $N$-$A$ scattering. 

To calculate $V_{np}\ket{\Psi_{d}^{(+)}}$ the ADWA expands the three-body wave function $\Psi^{(+)}_{d}(\textbf{R},\textbf{r})$ in a discrete set of states using the Weinberg eigenstates \cite{JT}, defined by
\begin{equation}
[-\epsilon_{d}-T_{r}-\alpha_{i}V_{np}]\phi_i(\textbf{r}) = 0, \qquad i=1,2,...,
\label{eigen}
\end{equation}
where the $\phi_i$ satisfy the orthonormality relation
\beq
\braket{\phi_i|V_{np}|\phi_j}=-\delta_{i,j}.
\eeqn{orthonormality}
The eigenvalue equation (\ref{eigen}) features a fixed deuteron energy $-\epsilon_{d}$ and $n$-$p$ kinetic energy operator $T_{r}$. The $\alpha_i$ increase monotonically, with $\phi_i$ possessing $i$ nodes within the range of $V_{np}$, such that $\phi_i$ becomes increasingly oscillatory. The $\Psi^{(+)}_{d}(\textbf{R},\textbf{r})$ is thus expanded in this basis
\beq
\Psi^{(+)}_{d}(\textbf{R},\textbf{r})=\sum^{\infty}_{i=1} \phi_{i}(\textbf{r})\chi_{i}^{(+)}(\textbf{R}),
\eeqn{weinbergexp}
where
\beq
\chi_{i}^{(+)}(\textbf{R})=-\braket{\phi_i|V_{np}|\Psi^{(+)}_d}.
\eeqn{weinbergchi}
In the ADWA the first Weinberg component of $\Psi^{(+)}_{d}$, which provides the dominant contribution to the $(d,p)$ stripping amplitude \cite{Pang}, is found by determining the distorted wave, $\chi^{(+)}_{d}$, which is the solution to the Schr\"odinger equation
\beq
(E_{d}-T_{R}-\braket{\phi_{1}\phi_A|U|\phi_{0}\phi_A})\chi^{(+)}_{d}(\ve{R})=0,
\eeqn{2beq}
where $T_{R}$ is the kinetic energy associated with the $n$-$p$ centre-of-mass coordinate $\ve{R}=(\ve{r}_{n}+\ve{r}_{p})/2$, and $\phi_{A}$ and $\phi_{0}$ are the target and deuteron ground state wave functions, respectively, while $\ket{\phi_{1}}$ is given by
\begin{equation}
\ket{\phi_{1}}=\dfrac{V_{np}\ket{\phi_{0}}}{\braket{\phi_{0}|V_{np}|\phi_{0}}}.
\end{equation}
The matrix element in $\braket{\phi_{1}\phi_{A}|U|\phi_{0}\phi_{A}}$ from Eq.(\ref{2beq}) implies integration over all internal degrees of freedom of $A$ and over $n$ and $p$ spin coordinates together with the relative $n$-$p$ coordinate $\ve{r} =\ve{r}_{n}-\ve{r}_{p}$. This leaves $\braket{\phi_{1}|U|\phi_{0}}$ to be an operator in the space of the  coordinate $\ve{R}$ and the triplet spin-space of the $n$ and $p$ \cite{Joh14}.

In Ref.\cite{Joh14} it was assumed that $\braket{\phi_{1}|U|\phi_{0}}\approx \braket{\phi_{1}|U^{(0)}|\phi_{0}}$ and it was shown that because of the short-range nature of $\phi_1$ the averaging procedure results in
\beq
\braket{\phi_{1}\phi_A|U^{(0)}|\phi_A\phi_{0}} \approx \sum_{N = n,p} 
\braket{\phi_1\phi_{A}|U_{NA}^{opt}(E_{\text{eff}})|\phi_0\phi_{A}},
\,\,\,\,\,\,\,\,\,\,\,\,\,
\eeqn{u0}
where $U_{NA}^{\text{opt}}$ is the optical model operator 
\beq
U_{NA}^{\text{opt}}(E_{\text{eff}})&&=v_{NA}\eol+v_{NA}&&\frac{Q_A}{E_{\text{eff}}-T_N-H_A-Q_Av_{NA}Q_{A}}v_{NA}.
\eeqn{UNA}
taken at energy
\beq
E_{\text{eff}}=\frac{1}{2}E_{d}+\frac{1}{2}\braket{T_{r}},
\eeqn{Eeff}
which differs from the commonly used value of half the deuteron incident energy by half the $n$-$p$ kinetic energy, $T_{r}$, averaged over the short range of the $n$-$p$ interactions:
\begin{equation}
\braket{T_{r}}=\braket{\phi_{1}|T_{r}|\phi_{0}}.
\end{equation}
This form (\ref{UNA}) for the optical model operator differs from that in Eq.(\ref{feshUNA}) only in the energy denominator, as now $U^{opt}_{NA}$ describes the two-body $N$-$A$ scattering the effective energy $E_{\text{eff}}$.

We want to show now that the same ideas allow us to recover some of the contributions from $U^{(1)}$. Approximating $U_{NA}$ in Eq.(\ref{u1}) by its leading value of $v_{NA}$ we obtain the contribution to the deuteron distorted potential given by
\beq
\begin{aligned}
\bra{\phi_{1} \phi_{A},\textbf{R}}U^{(1)}&\ket{\phi_{A} \phi_{0},\textbf{R}'} \approx \\
&\bra{\phi_{1}\phi_{A},\textbf{R}}v_{nA}\dfrac{Q_{A}}{e}v_{pA}\ket{\phi_{A}\phi_{0},\textbf{R}'} \\
+&\bra{\phi_{1}\phi_{A},\textbf{R}}v_{pA}\dfrac{Q_{A}}{e}v_{nA}\ket{\phi_{A}\phi_{0},\textbf{R}'} \\
&\equiv U^{(1)}_{np}(\ve{R},\ve{R}') + U^{(1)}_{pn}(\ve{R},\ve{R}').
\end{aligned}
\eeqn
This potential is nonlocal in space $\ve{R}$. To estimate its magnitude we consider the case of local, spin independent interactions $v_{NA}$, and ignore Coulomb contributions.  We rewrite $ U^{(1)}_{np}(\ve{R},\ve{R}')$ as
\beq
 U^{(1)}_{np}(\ve{R},\ve{R}')
&=&\int d\xi_{A} d\ve{r} \, \phi_{1}^{*}(\ve{r})\phi_{A}^{*}(\xi_{A})
v_{nA}(\ve{R}-\tfrac{\ve{r}}{2},\xi_{A}) 
\eol
&\times& 
{\tilde \Psi}_p(\ve{r}_n,\ve{r}_p,\ve{R}',\xi_{A}),
\eeqn{u1np}
where $\ve{r}_n= \ve{R}-\tfrac{\ve{r}}{2}$, $\ve{r}_p=\ve{R}+\tfrac{\ve{r}}{2}$ and 

\beq
\begin{aligned}
{\tilde \Psi}_{p}(\ve{r}_n,\ve{r}_p,\ve{R}',\xi_{A})&= \\
&\bra{\ve{r}_{n},\ve{r}_p,\xi_{A}}\frac{Q_A}{e}v_{pA}\ket{\phi_0 \phi_A,\textbf{R}'}. 
\end{aligned}
\eeqn
Given the short range of $\phi_1$ we can replace $\ve{r}_p$ and $\ve{r}_n$ in ${\tilde \Psi}_p$ by $\ve{R}$, and $\ve{r}$ in $v_{nA}$ by zero in Eq.(\ref{u1np}). Then the expression for $U^{(1)}_{np}$ becomes very similar to the one for $U^{(0)}_{pp}$, arising from averaging the second term on the right in $U_{pA}$ in Eq.(\ref{UNNA}) and treated using the same approximation. The only difference is the presence of $v_{nA}(\ve{R},\xi_A)$ instead of $v_{pA}(\ve{R},\xi_A)$  in the integrand of the right-hand side of (\ref{u1np}). If we further assume that the interaction of the proton $p$ and neutron $n$ with the nucleons of target $A$ are the same we obtain
\beq
\la \phi_{1}\phi_{A}|U^{(0)}&+&U^{(1)}|\phi_{A}\phi_{0}\ra \approx \sum_{N=n,p}\la \phi_1 \phi_A | v_{NA}
\eol 
&+&2v_{NA}\dfrac{Q_{A}}{e-Q_{A}v_{NA}Q_{A}}v_{NA}|\phi_0 \phi_A\ra. \,\,\,\,\,\,\,\,\,\,
\eeqn{}
Further reasoning along the lines in \cite{Joh14} leads to conclusion that $e$ could be substituted by $E_{\text{eff}}+i\epsilon-T_{N}-H_{A}$, so that
\beq
\la \phi_{1}\phi_{A}|U^{(0)}+U^{(1)}|\phi_{A}\phi_{0}\ra \approx \sum_{N=n,p}&\la&\phi_{1}\phi_{A}|v_{NA}
\eol 
+2v_{NA}\dfrac{Q_{A}}{E_{\text{eff}}-T_N-H_A-Q_Av_{NA}Q_{A}}&v&_{NA}Q_{A}|\phi_0 \phi_A\ra. \,\,\,\,\,\,\,\,\,\,
\eeqn{approximate}
When compared to our expression for $\braket{\phi_{1}\phi_A|U^{(0)}|\phi_A\phi_{0}}$ in Eq.(\ref{u0}) we can see that $\braket{\phi_{1}\phi_{A}|U^{(0)}+U^{(1)}|\phi_{A}\phi_{0}}$ includes a factor of two in the term corresponding to the second term in Eq.(\ref{UNNA}). As such, it is possible to rewrite $\braket{\phi_{1}\phi_{A}|U^{(0)}+U^{(1)}|\phi_{A}\phi_{0}}$ in terms of $\braket{\phi_{1}\phi_A|U^{(0)}|\phi_A\phi_{0}}$ as
\beq
\la \phi_{1}\phi_{A}|U^{(0)}&+&U^{(1)}|\phi_{A}\phi_{0}\ra \approx  2\la \phi_1 \phi_A | U^{(0)}| \phi_0 \phi_A \ra
\eol
&-& \sum_{N=n,p} \la \phi_1 \phi_A | v_{NA}| \phi_0 \phi_A \ra.
\eeqn{}
\section{Connection with Dispersive Optical Model.}

It has been shown in the previous section that I3B terms arising from $U^{(1)}$ can be accounted for in the ADWA to first order by doubling the adiabatic deuteron optical potential and subtracting from it the Johnson-Tandy potential calculated from nucleon-target (real) folding potentials. The nucleon optical potentials, $V^{opt}_{NA}(E) \equiv \la \phi_A|U^{opt}_{NA}|\phi_A\ra$, needed to construct the deuteron adiabatic distorting potential should be nonlocal, energy-dependent and complex since they arise due to  projecting out the space of excited states of the target \cite{Fesh}. It is also known that, due to causality, optical potentials fulfill a dispersion relation \cite{mahauxsartor}. This means that the optical potential consists of two terms,
\beq
\begin{aligned}
V^{opt}_{NA}(E) &= V_{NA}^{HF} +\Delta V_{NA}^{dyn}(E),
\end{aligned}
\eeqn{VDOMNA}
one of which, $V_{NA}^{ HF}$, is a real energy-independent potential and the other, $\Delta V_{NA}^{dyn}(E)$, is generated dynamically though coupling to inelastic channels and is energy-dependent \cite{NLDOM}. This complex term has an imaginary part $W_{NA}(E)$ and a real part that is related to $W_{NA}(E)$ by the dispersive relation, so that
\beq
\begin{aligned}
\Delta V_{NA}^{dyn}(E)&=iW_{NA}(E)+\dfrac{\mathcal{P}}{\pi}\int^{\infty}_{-\infty} dE'\dfrac{W_{NA}(E')}{E-E'}.\end{aligned}
\eeqn{deltaVNA}
We will identify the real, energy independent, folding term in Eq.(\ref{approximate}), $\la \phi_A | v_{NA}| \phi_A \ra$, with $V^{HF}_{NA}$, ignoring the fact that the Feshbach formalism used here does not carry the exchange terms needed to generate $V^{HF}_{NA}$, and identify the second term in Eq.(\ref{approximate}) with the dynamical term $\Delta V_{NA}^{dyn}(E)$ multiplied by two. Then,
\beq 
\braket{\phi_A | U^{(0)}+U^{(1)}| \phi_A} &=& V^{HF}_{nA} + 2\Delta V_{nA}^{dyn}(E), \eol
&+& V^{HF}_{pA} + 2\Delta V_{pA}^{dyn}(E).  \eol
\eeqn{u0u1}
This equation provides an approximate practical approach to estimating the effect of the I3B terms if both the parts $V^{HF}_{NA}$ and $\Delta V_{NA}^{dyn}(E)$ are known. These potentials should be used at an energy $E = E_{\rm eff}$ defined in Eq. (\ref{Eeff}).
Recently, a phenomenological NLDOM parameterisation has been proposed to forge the link between nuclear structure and reactions \cite{NLDOM} for proton and neutron scattering from $^{40}$Ca. Below, we will use this potential for the  $^{40}$Ca$(d,p)^{41}$Ca calculations, employing the updated NLDOM parameters from Ref. \cite{Wal16}.

\section{nonlocal scattering.}

The ADWA calculations require knowledge of the distorted waves $\chi_d$ and $\chi_p$ in the entrance and exit channels. For nonlocal optical potentials they are found from equations \cite{Gregg}:
\beq
(T_{d}^{(L)}&+&U_{c}(R_d)-E_{d})\chi^{J}_{L'L}(R_d)= 
\eol
&-&\sum_{L''}\int^{\infty}_{0}dR'_d \,R_dR'_d \,\mathcal{U}^{J}_{L'L''}(R_d,R'_d)\, \chi^{J}_{L'}(R'), \,\,\,\,\,\,\,\,\,\,\,\,\,\,\,\,
\eeqn{chid}
and 
\beq
(T_{p}^{(L)}&+&U_{c}(R_{p})-E_{p})\chi^{J}_{L}(R_{p})= 
\eol
&-&\int^{\infty}_{0}dR_{p}' \, R_{p}R_{p}'\,\mathcal{U}^{J}_{L}(R_{p},R_{p}')\, \chi^{J}_{L}(R_{p}'), \,\,\,\,\,\,\,\,\,\,\,\,\,\,\,\,\,\,\,\,
\eeqn{chip}
where $T_\alpha^{(L)} $ is the kinetic energy operator in a partial wave with orbital angular momentum $L$ in the channel $\alpha$, characterized by energy $E_{\alpha}$ and reduced mass $\mu_{\alpha}$:
\begin{equation}
T_\alpha^{(L)} = -\frac{\hbar^2}{2\mu_\alpha}\left[\frac{d^2}{dR^2_\alpha}-\frac{L(L+1)}{R^2_\alpha}\right],  \text{ }\text{ }\text{ }\text{ } \alpha=d,p.
\end{equation}
Equations (\ref{chid}) and (\ref{chip}) contain the Coulomb interaction $U_{c}$ and nonlocal deuteron-channel  ${\cal U}^{J}_{L'L''}(R_d,R'_d)$ and 
proton-channel ${\cal U}^{J}_{L} (R_p,R_p')$ potential kernels that depend on channel coordinates $R_{\alpha}$ and $R_{\alpha}'$ (note that $\ve{R}_d \equiv \ve{R})$. In these equations $J$ is the total angular momentum in the scattering channels. We neglect spin-orbit interaction in the present work. The correct description of its effects within the ADWA requires spin-dependent tensor terms \cite{Joh15}, for which no numerical implementations are yet available.

The nucleon NLDOM potential used in this work consists of seven terms,
\beq
{\cal U}^{\rm NLDOM}(\ve{R},\ve{R'}) = \sum^7_{i=1} U_i\left(\dfrac{|\ve{R}+\ve{R}'|}{2}\right)H_i(s),
\eeqn{NLDOM}
where $\ve{s} = \ve{R} - \ve{R}'$, each described by its own nonlocality range $\beta_i$ in the nonlocal factor
 \begin{equation}
H_i(s)=\dfrac{\text{exp}(-s^{2} /\beta_i^{2})}{\pi^{\frac{3}{2}}\beta_i^{3}}.
\end{equation}
Details of nonlocal kernel calculations for one nonlocality range and the $s$-wave deuteron only are given in \cite{Tit16}, while the generalisation to a realistic deuteron wave function  that includes the deuteron $d$-state is available in \cite{Gregg}. Nonlocal kernels with several nonlocality ranges are just the sums of kernels calculated with one nonlocality range. 

It has been shown in \cite{Bai16} that including the deuteron $d$-wave component in ADWA leads to enhanced sensitivity of $(d,p)$ cross sections to high $n$-$p$ momenta, which is an artifact of the adiabatic approach \cite{Gom18,Del18}.  It has also been shown in \cite{Gom18} that ADWA and beyond-ADWA calculations differ less when only a deuteron  $s$-state is included. For this reason we use the Hulth\'en model for $\phi_{1}$ and $\phi_{0}$, which does not contain any deuteron $d$-state \cite{Hulthen}, in all our calculations of deuteron-channel nonlocal kernels.

We have generated the deuteron channel NLDOM kernels for each nonlocality range and then summed them.  To test our procedure, we calculated the kernels in the leading-order approximation $U_i\left(|\ve{R}+\ve{R}'|/2\right) \approx U_i\left(\ve{R}\right)$. For one nonlocality range, this approximation gives a very similar result to the calculations performed with a local-equivalent potential $U_{loc}$ obtained as a solution of a transcendental equation (see \cite{Gregg}). In the case of NLDOM, $U_{loc}$ is the solution of the generalised transcendental equation \cite{Wal16}
\beq
U_{loc} = \sum^7_{i=1} U_i \exp\left( \frac{\mu_d \beta^2_i}{2 \hbar^2}(E_d -U_c - U_{loc})\right).
\eeqn{uloc}
We have checked that just as in \cite{Gregg}, the calculations with $U_{loc}$ differ from the leading-order calculations at the cross section peak by 1$\%$ at 11.8 and 20 MeV respectively, but differ by up to 5$\%$ for 56 MeV. Exact solutions of the nonlocal Eq.(\ref{chid}) reduce the leading-order $(d,p)$ cross sections in the main peak by no more than 10$\%$ for all the deuteron energies, with the largest differences due to a small change in the location of the peak. This holds for all the optical models in the proton channel that were considered here. In the p-channel, exact solutions of nonlocal Eq.(\ref{chip}) again reduce the $(d,p)$ cross sections in the main peak by no more than 10$\%$ for all investigated energies.

It was also found that the difference between exact and transcendental methods are smaller when I3B terms are accounted for, with said differences in cross sections reduced in both the $d$ and $p$-channel to no more than 5$\%$ at the main peak for all investigated energies. This change is because the nonlocal wave functions obtained from Eqs.(\ref{chid}) and (\ref{chip}) are smaller in the nuclear interior than those obtained from the local Schr\"odinger equation used in conjunction with Eq.(\ref{uloc}). This difference is given by the Perey factor \cite{Perey}
\beq
f(r) = \text{exp}\left(\dfrac{\mu_{d}\beta^{2}}{4\hbar^{2}}U_{loc}\right),
\eeqn{pereyfactor}
which is equal to unity outside the nucleus. With increased absorption, the contribution from the nuclear interior becomes less important, so applying the Perey factor does not have the same effect. This leads to results closer to those found using transcendental methods, reducing the difference between the nonlocal and local solutions.

\section{The $^{40}{\rm Ca}(d,p)^{41}{\rm Ca}$ results.}

We have carried out numerical calculations of the $^{40}$Ca$(d,p)^{41}$Ca reaction using the  NLDOM nucleon potential in the entrance deuteron channel evaluated at $E_{\text{eff}}$ according to Eq.(\ref{Eeff}) and \cite{Joh14}. The shift of $\la T_r\ra/2 = 57$ MeV was applied, consistent with the Hulth\'en model, which does not contain high relative $n$-$p$  momenta. In the exit channel the NLDOM was used at the actual proton energy.
The calculations have been carried out at $E_{d}$ = 11.8, 20 and 56 MeV. These cross sections have been measured in Refs. \cite{40Ca12MeV,40Ca20MeV,40Ca56MeV1,40Ca56MeV2} and we note that at 56 MeV the two measured data sets differs significantly. In all of the nonlocal ADWA calculations presented, the exact solutions of Eqs.(\ref{chid}) and (\ref{chip}) for deuteron and proton distorted waves in the entrance and exit channels are read into the transfer reactions code TWOFNR \cite{TWOFNR} and the transition amplitude is calculated in the zero-range approximation using a standard value of $D_0 = -126.15$ MeV fm$^{3/2}$. 

The overlap integral between the $^{41}$Ca and $^{40}$Ca ground state wave functions was taken from Ref. \cite{Wal16}, where the exact NLDOM overlap function was approximated, with good accuracy, by the single-particle wave function calculated in a Wood-Saxon potential well  with the radius $r_0$ = 1.252 fm, diffuseness $a = 0.718$ fm and spin-orbit depth $V_{s.o.}= 6.25$ MeV. This single-particle wave function has been multiplied by the square root of the NLDOM spectroscopic factor, $S = 0.73$. 

The results of the calculations are plotted in Fig.1. It was already noticed in \cite{Wal16} that the NLDOM overestimates significantly the  $^{40}$Ca$(d,p)^{40}$Ca cross section at 11.8 MeV. The cross sections at the two other energies we investigate are also overestimated. Moreover, the shape of neither of the two available 56 MeV data sets are reproduced. Including first-order I3B terms, by doubling the dynamical real and imaginary NLDOM parts, decreases the cross sections due to increased absorption by a factor shown in Table 1, bringing them to much better agreement with experimental data at 11.8 and 20 MeV, with 56 MeV data remaining difficult to reproduce.

\newcolumntype{C}[1]{>{\centering\arraybackslash}p{#1}}
\begin{table}[h!]
\begin{tabular}{C{1.8cm}C{1.7cm}C{1.7cm}C{1.7cm}}
\hline 
\hline 
& \multicolumn{3}{c}{Ratio of $\sigma^{peak}_{I3B}/\sigma^{peak}$} \\ 
\hhline{~|---|}
$E_{d}$ (MeV)& NLDOM & GR & GRZ \\ 
\hline
11.8 & 0.725 & 0.729 & 0.630 \\ 
20 & 0.691 & 0.706 & 0.621 \\ 
56 & 0.821 & 0.842 & 0.786 \\
\hline 
\hline 
\end{tabular} 
\caption{The factor by which I3B terms change the size of the cross section peaks. Presented as the ratio between the peak maximum when including first-order I3B, $\sigma^{peak}_{I3B}$, and without them, $\sigma^{peak}$. Peaks from Figs.1 and 2 for each investigated optical potential at $E_{d}$ = 11.8, 20 and 56 MeV.}
\end{table}

To investigate the role of the dynamical real part, we present calculations where only the imaginary part of the NLDOM is doubled. Using Re$\left[\Delta V^{dyn}_{NA}\right]$ instead of 2Re$\left[\Delta V^{dyn}_{NA}\right]$ makes no notable difference to the calculated cross sections (see Fig.1). We have found that this effect is only present when working with a doubled imaginary part, and the large absorption this implies. When the absorption is increased in such a manner, the effect of altering the dynamical real part of the potential becomes less notable, as it corresponds to a much smaller proportional change to the absolute optical potential than it would for an unmodified imaginary component. This could be a useful insight as several phenomenological systematics that do not include dynamical dispersive corrections are available \cite{GR,GRZ,MSU,Jaghoub}, and our findings suggest that omitting a real dispersive term may not be significant in $(d,p)$ calculations when I3B terms are present. 

It is important to note that this statement is also true for cross sections produced from deuteron-target potentials found using the transcendental method Eq.(27) and for its linear approximation, making it clear that this behaviour is a result of the potential itself, rather than the method used to generate results.

Further calculations were carried out with phenomenological nonlocal optical potentials, the energy independent Giannini-Ricco (GR) \cite{GR} and energy dependent Giannini-Ricco-Zucchiatti (GRZ) \cite{GRZ}. Energy dependent $n$-$A$ and $p$-$A$ potentials based on the Tian, Pang, Ma \cite{TPM} optical potentials have become available recently \cite{MSU,Jaghoub}. Unfortunately these do not cover the nucleon energy range based on Eq.(10) that is needed here. The GR and GRZ potentials include only the imaginary term $W_{NA}$ from $\Delta V^{dyn}_{NA}$, so accounting for first-order I3B terms would imply doubling the well depths of the imaginary part only, with no changes being made to the real part of each potential. These imaginary terms feature a surface term only. The calculations with these two potentials are presented in Fig.2 and they show, qualitatively, the same result found when using NLDOM: that including first-order I3B terms decreases the cross sections by factors similar to those obtained with the NLDOM, shown in Table 1 for each energy. The 11.8 MeV data are better reproduced with the GR potential while the 20 MeV data favour the GRZ calculations. The case of 56 MeV remains inconclusive. There are large discrepancies between two sets of measurements in the literature \cite{40Ca56MeV1,40Ca56MeV2} and resolving them experimentally is an important task.


\begin{figure}[t!]
\includegraphics[scale=0.3]{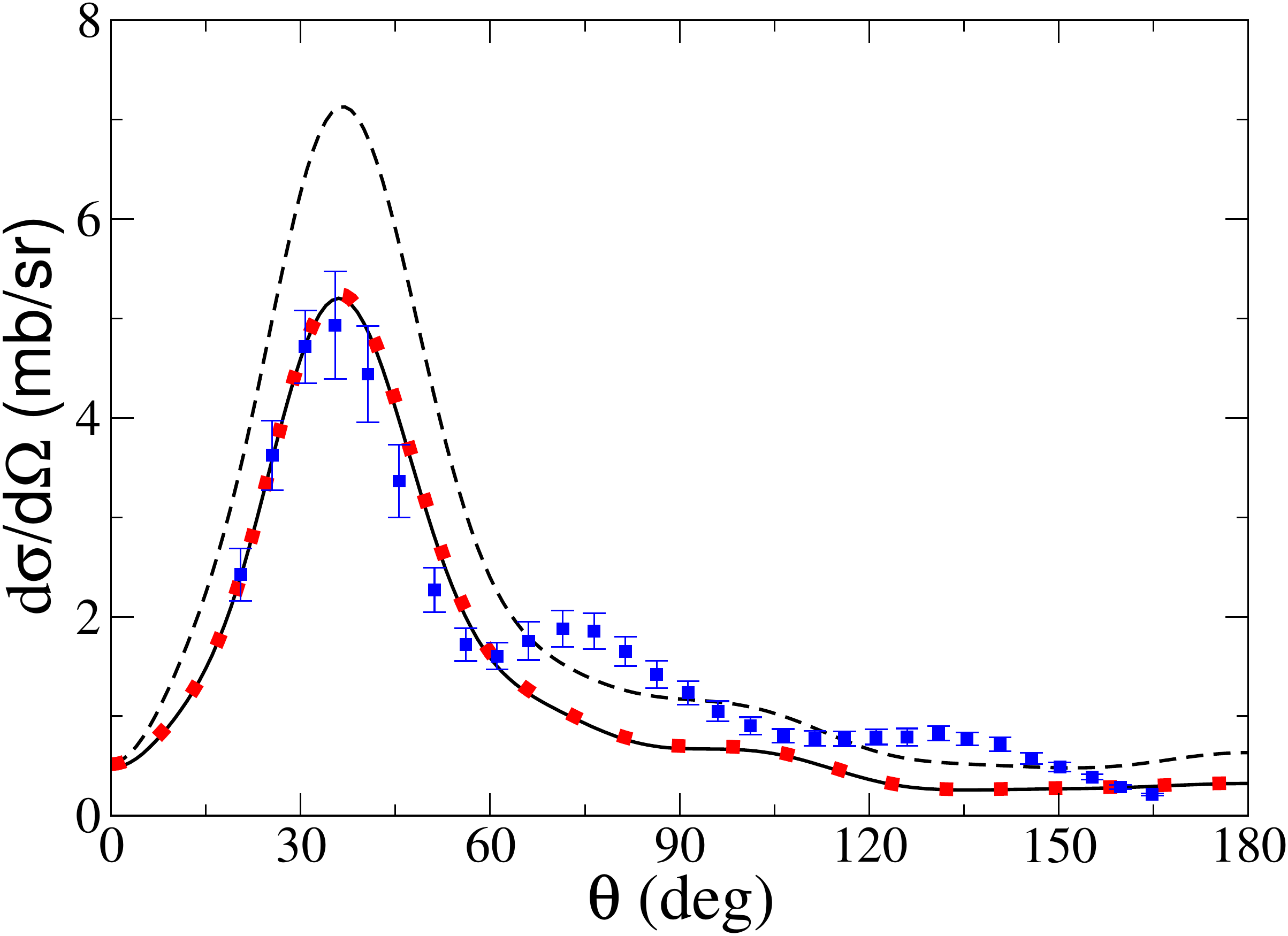}
\includegraphics[scale=0.3]{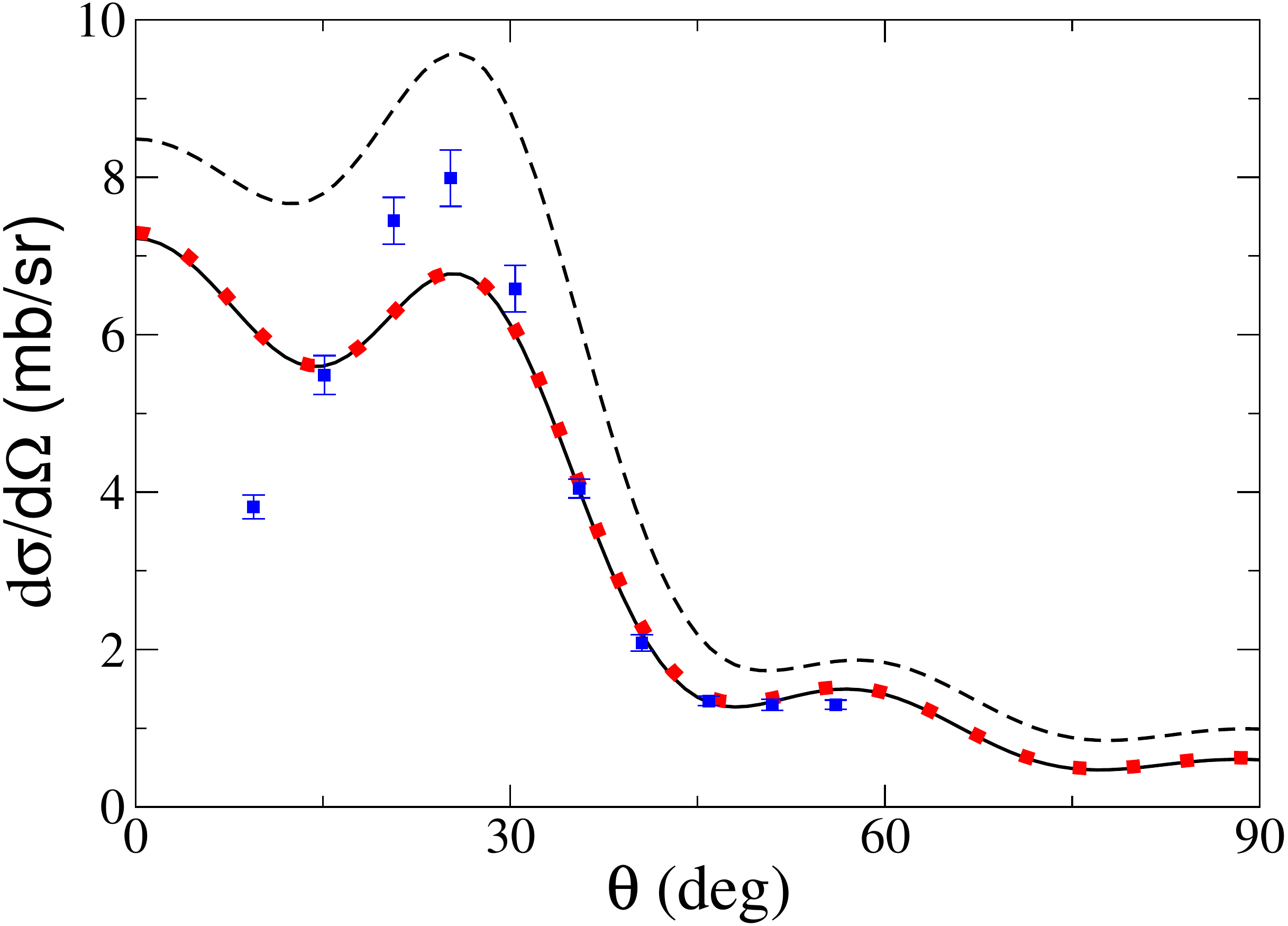}
\includegraphics[scale=0.3]{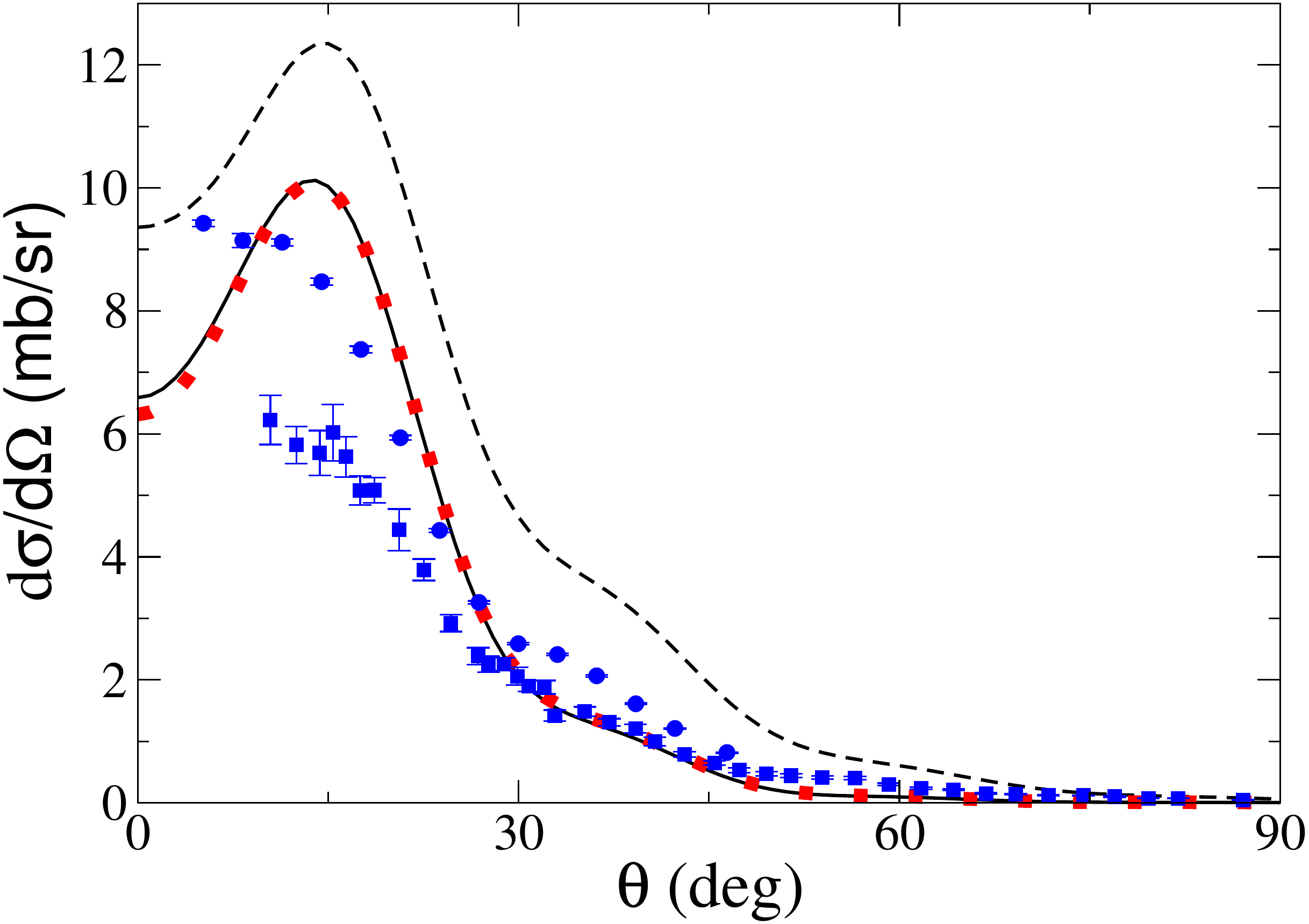}
\caption{Comparison of $^{40}$Ca$(d,p)^{40}$Ca cross sections for the ground state at $E_{d}$ = 11.8 MeV (top), 20 MeV (middle) and 56 MeV (bottom) using the DOM optical potentials. Cross sections are found with (solid lines) and without (dashed lines) I3B effects. We also present results for when Re$\left[\Delta V^{dyn}_{NA}\right]$ is left unmodified (dotted lines). Experimental data from \cite{40Ca12MeV,40Ca20MeV,40Ca56MeV1,40Ca56MeV2}.}
\end{figure}

\begin{figure}[t!]
\includegraphics[scale=0.3]{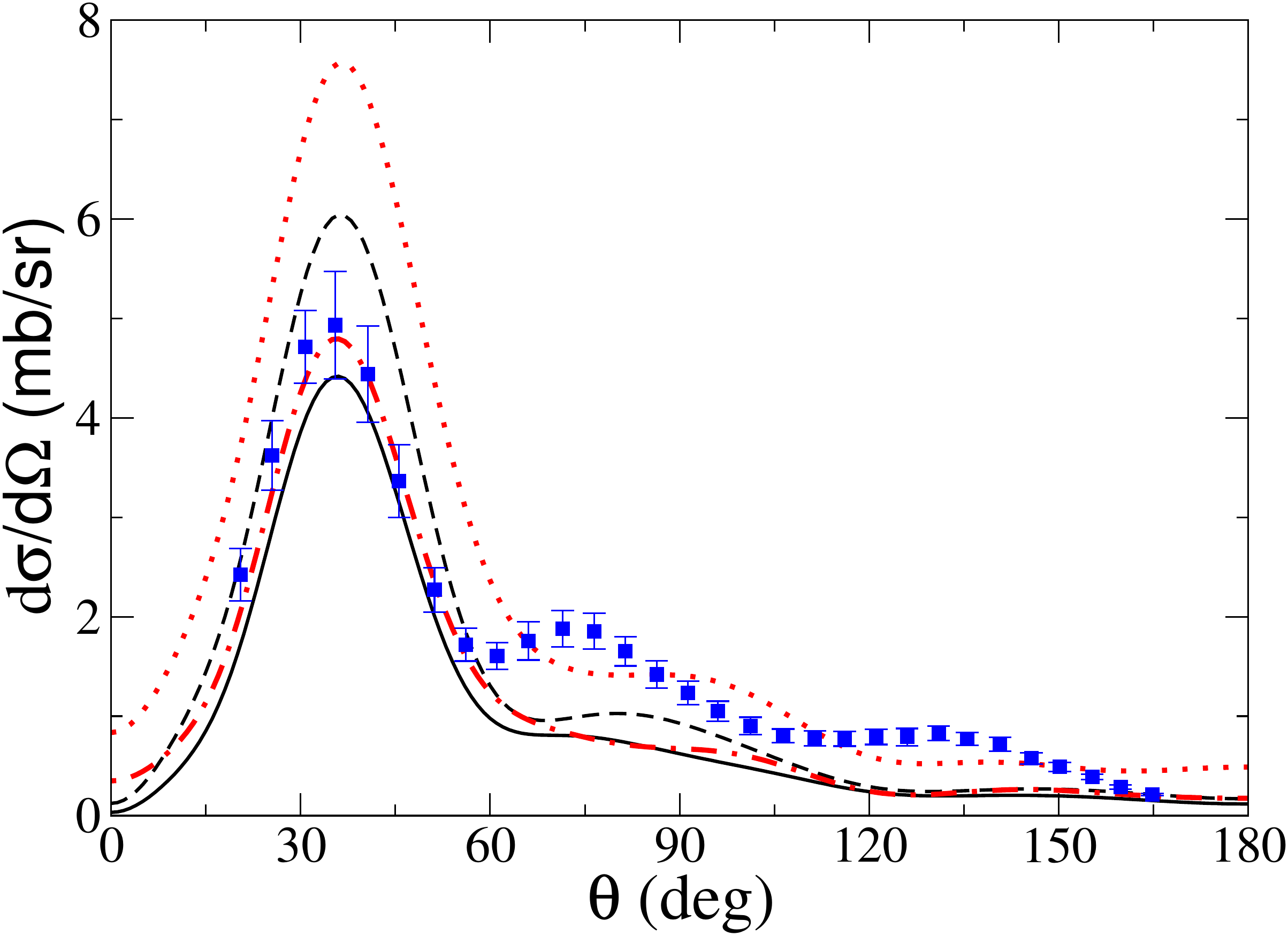}
\includegraphics[scale=0.3]{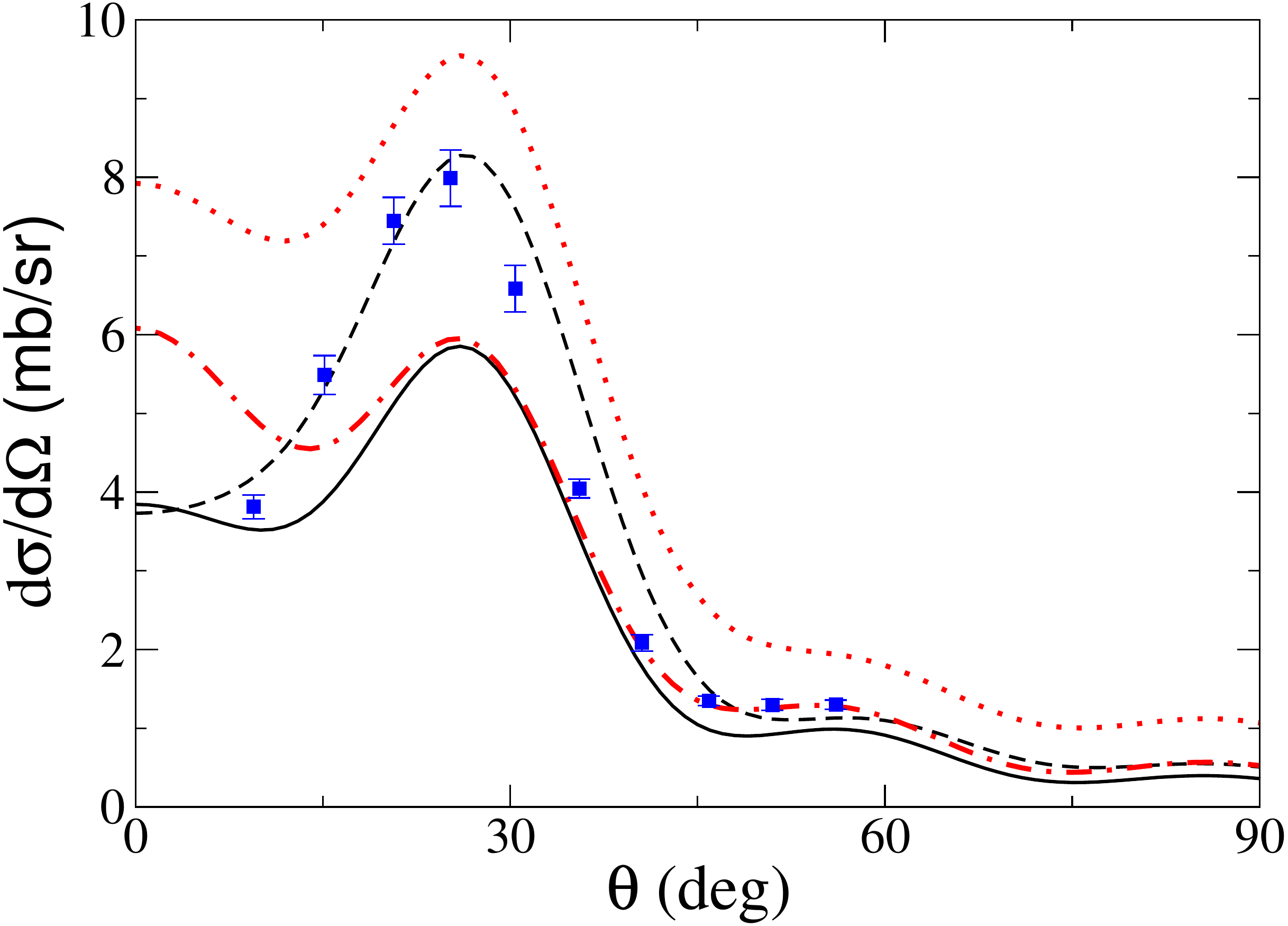}
\includegraphics[scale=0.3]{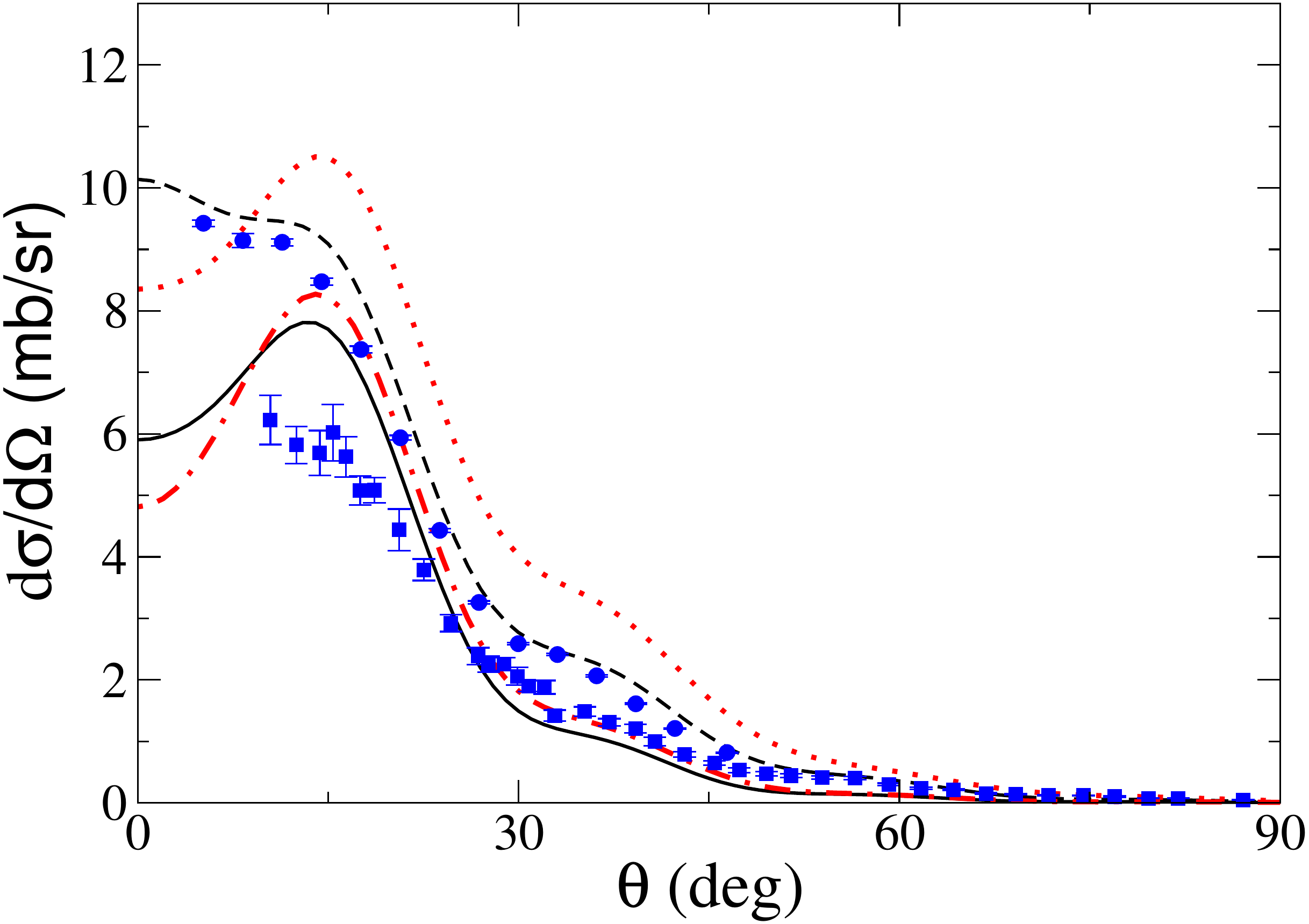}
\caption{Comparison of $^{40}$Ca$(d,p)^{40}$Ca cross sections for the ground state at $E_{d}$ = 11.8 MeV (top), 20 MeV (middle) and 56 MeV (bottom) using the GR potential with (solid lines) and without (dashed lines) I3B effects, along with cross sections found with the GRZ optical potential with (dot-dashed lines) and without (dotted lines) I3B effects. Experimental data from \cite{40Ca12MeV,40Ca20MeV,40Ca56MeV1,40Ca56MeV2}.}
\end{figure}



\section{Summary and Conclusions.}
We have considered new terms in the optical potential in the incident channel in the  ADWA approximation for $A(d,p)B$ reactions.
These terms arise because  target excitations  in the $A+n+p$ breakup channels are coupled  to different nucleons in the deuteron.
These couplings are usually neglected when $U$ is approximated as the sum of $n$-$A$ and $p$-$A$ optical model operators. Treating the new terms in the ADWA approximation  leads to an effective operator $U$ in which the first-order I3B terms double the dynamical excitation contributions to the $n$-$A$ and $p$-$A$ optical potentials.

Numerical calculations, performed for the $^{40}$Ca(d,p)$^{41}$Ca reaction with NLDOM, GR and GRZ potentials, have shown that these I3B terms decrease the ADWA cross sections by 20-40$\%$, depending on the deuteron energy, bringing the cross sections closer to available experimental data. It was found that with stronger imaginary parts the impact of modifying the dynamical real part becomes insignificant. This suggests that other (non-dispersive)  nonlocal optical potentials could be used if I3B terms are taken into account by simply doubling their imaginary parts in a standard ADWA calculation. Given that the NLDOM is available only for  nucleon scattering from $^{40}$Ca \cite{NLDOM} and $^{48}$Ca \cite{48Ca} this finding could be useful for applications to all other nuclei, as it would allow for the use of existing global nucleon optical potentials without dispersive terms.

Finally, our estimations suggest that I3B effects could play an important role in forming both the shape and absolute magnitude of $(d,p)$ differential cross sections. This could have important consequences for extracting spectroscopic information from $(d,p)$ experiments. It is important to extend the investigation of I3B force effects in the $d+A$ system beyond the approximate methods used here.

\section*{Acknowledgements.}
We are grateful to Professor J.A. Tostevin for providing codes that treat the d-channel and p-channel using the exact methods described in Section IV, making modifications to the p-channel channel code to allow for the multiple nonlocality ranges required to treat NLDOM as well as support with the use of these tools. This work was supported by the United Kingdom Science and Technology Facilities Council (STFC) under Grant No. ST/L005743/1.


\begin{thebibliography}{1}
\bibitem{JT} R.C. Johnson and P.C. Tandy, Nucl. Phys. A \textbf{235}, 56 (1974).
\bibitem{Feshbach} H. Feshbach, Ann. Phys. \textbf{5}, 357 (1958).
\bibitem{Joh14} R.C. Johnson and N.K. Timofeyuk, Phys. Rev. C \textbf{89}, 024605 (2014).
\bibitem{Wal16} S.J. Waldecker and N.K. Timofeyuk, Phys. Rev. C \textbf{94}, 034609 (2016).
\bibitem{Gom15} M.Gomez-Ramos, A.M.Moro, J.Gomez-Camacho, I.J.Thompson, Phys.Rev. C \textbf{92}, 014613 (2015).
\bibitem{Gom17} M.Gomez-Ramos and A.M.Moro,  Phys.Rev. C \textbf{95}, 044612 (2017).
\bibitem{Del15} A.Deltuva, Phys.Rev. C \textbf{91}, 024607 (2015).
\bibitem{Del16} A.Deltuva, A.Ross, E.Norvaisas, F.M.Nunes, Phys.Rev. C \textbf{94}, 044613 (2016).
\bibitem{Del17} A.Deltuva, D.Jurciukonis, E.Norvaisas, Phys. Lett. B \textbf{769}, 202 (2017).
\bibitem{Feshbach2} H. Feshbach, Ann. Phys. \textbf{19}, 287 (1962).
\bibitem{Pang} D. Y. Pang, N. K. Timofeyuk, R. C. Johnson, and J. A. Tostevin, Phys. Rev. C \textbf{87}, 064613 (2013).
\bibitem{Fesh} H. Feshbach, Ann. Rev. Nucl. Sci. \textbf{8}, 49 (1958).
\bibitem{mahauxsartor} C. Mahaux, R. Sartor, Nucl. Phys. A, \textbf{484}, 205 (1988).
\bibitem{NLDOM} M. H. Mahzoon, R. J. Charity, W. H. Dickhoff, H. Dussan, S. J. Waldecker, Phys. Rev. Lett. \textbf{112}, 162503 (2014).
\bibitem{Gregg} G. W. Bailey, N. K. Timofeyuk, J. A. Tostevin, Phys. Rev. C \textbf{95}, 024603 (2017).
\bibitem{Joh15} R.C. Johnson,  Phys.Rev. C \textbf{91}, 054604 (2015).
\bibitem{Tit16} L.J.Titus, F.M.Nunes, G.Potel, Phys.Rev. C \textbf{93}, 014604 (2016).
\bibitem{Bai16}  G. W. Bailey, N. K. Timofeyuk, J. A. Tostevin, Phys. Rev. Lett. \textbf{117}, 162502 (2016).
\bibitem{Del18} A. Deltuva,  Phys.Rev. C \textbf{98}, 021603 (2018).
\bibitem{Gom18} M. Gomez-Ramos and N.K. Timofeyuk, Phys. Rev. C \textbf{98}, 011601 (2018).
\bibitem{Hulthen} Y. Yamaguchi,Phys. Rev. \textbf{95}, 1628 (1954).
\bibitem{Perey} F. Perey, \textit{Direct Interactions and Nuclear Reaction Mechanisms} (Gordon and Breach, New York, 1963), p. 125.
\bibitem{40Ca12MeV} U. Schmidt-Rohr, R. Stock, and P. Turek, Nucl. Phys. \textbf{53}, 77 (1964).
\bibitem{40Ca20MeV} F.J. Eckle, et al, Nuc. Phys. A \textbf{506}, 159 (1990).
\bibitem{40Ca56MeV1} K.Hatanaka, et al, Nuc. Phys. A \textbf{419}, 530 (1984).
\bibitem{40Ca56MeV2} Y.Uozumi, et al, Phys. Rev. C \textbf{50}, 263 (1994).
\bibitem{TWOFNR} J.A. Tostevin, University of Surrey version of the code {\sc twofnr} (of M. Toyama, M. Igarashi and N. Kishida) and code {\sc front} (private communication).
\bibitem{GR} M. M. Giannini, G. Ricco, Ann. Phys. \textbf{102}, 458 (1976).
\bibitem{GRZ} M. M. Giannini, G. Ricco, A. Zucchiatti, Ann. Phys. \textbf{124}, 208 (1980).
\bibitem{MSU} A.E. Lovell, P.-L. Bacq, P.Capel, F.M. Nunes and L.J. Titus, Phys. Rev. C \textbf{96}, 051601 (2017).
\bibitem{Jaghoub} M.I. Jaghoub, A.E. Lovell and F. M. Nunes, Phys. Rev. C \textbf{98}, 024609 (2018).
\bibitem{TPM} Y. Tian, D.-Y. Pang, and Z.-Y. Ma, Int. J. Mod. Phys. E \textbf{24}, 1550006 (2015).
\bibitem{48Ca} M. H. Mahzoon, PhD thesis, Washington University, 2015.

\end{thebibliography}
\end{document}